\documentclass[12pt]{iopart}
\usepackage{epsfig}
\usepackage{iopams}
\usepackage{epsfig} 
\usepackage{graphicx}
\usepackage{graphics}
\usepackage{subfigure}
\usepackage{verbatim} 
\newcommand{\eq}{\begin{eqnarray}} 
\newcommand{\en}{\end{eqnarray}} 
\newcommand{\ra}{\rangle}

\def\L2{\Lambda^2}

\def\US{\Upsilon}
\def\zb{Z_b^+}
\def\zbp{Z_b^{'+}}

\begin{document}
\title{Decays of $\zb$ and $\zbp$ as hadronic molecules\\}  
\author{Yubing Dong$^{1,2,3}$, 
        Amand  Faessler$^3$,   
        Thomas Gutsche$^3$, \\
        Valery E. Lyubovitskij$^3$\footnote{On leave of absence
        from Department of Physics, Tomsk State University,
        634050 Tomsk, Russia}
\vspace*{1.2\baselineskip}}
\address{
$^1$ Institute of High Energy Physics, Beijing 100049, P. R. China\\
 
$^2$ Theoretical Physics Center for Science Facilities (TPCSF), CAS, 
Beijing 100049, P. R. China\\

$^3$ Institut f\"ur~Theoretische Physik, Universit\"at T\"ubingen,\\ 
Kepler Center for Astro and Particle Physics, \\
Auf der Morgenstelle 14, D--72076 T\"ubingen, Germany\\}

\ead{
dongyb@ihep.ac.cn, 
amand.faessler@uni-tuebingen.de,\\
thomas.gutsche@uni-tuebingen.de,
valeri.lyubovitskij@uni-tuebingen.de} 

\date{\today}

\begin{abstract} 
The two newly observed hidden-bottom mesons $\zb(10610)$ and $\zbp(10650)$ 
with quantum numbers $J^P = 1^+$ are considered as hadronic molecules
composed of $\bar{B}B^*$ and $\bar{B}^*B^*$, respectively.
We give predictions for the widths of the strong two-body decays 
$\zb \to \US(nS)+\pi^+$ and $\zbp \to \US(nS)+\pi^+$ 
in a phenomenological Lagrangian approach. 
\end{abstract}

\vskip 1cm

\noindent {\it PACS:}
13.25.Gv, 13.30.Eg, 14.40.Rt, 36.10.Gv 

\noindent {\it Keywords:} 
bottom mesons, hadronic molecules, strong decays 

\maketitle

\section{Introduction}

Recently, two hidden-bottom charged meson resonances were observed 
by the Belle Collaboration~\cite{Adachi:2011} as two narrow resonance 
structures in the $\pi^{\pm}\US(nS)$ ($n=1,2,3$) and $\pi^{\pm}h_b(mP)$ 
$(m=1,2)$ mass spectra. 
They are produced in association with a single 
charged pion in $\US(5S)$ decays with the following values of 
mass and width: 
$M[\zb(1061)]=10608.4\pm 2.0$~MeV, 
$\Gamma[\zb(10610)]=15.6\pm 2.5$~MeV,  
$M[\zbp(10650)]=10653.2\pm 1.5$~ MeV, 
$\Gamma[\zbp(10650)]=14.4\pm 3.2$~MeV. 
Analyses of the charged pion angular distributions favor
the $I^G(J^P)=1^+(1^+)$ quantum numbers of the $Z$-states~\cite{Adachi:2011}. 

Theoretical structure assignments for these hidden-bottom 
meson resonances were proposed immediately after their 
observation~\cite{Voloshin:2011qa}-\cite{Richard:2011}, mainly based 
on molecular~\cite{Voloshin:2011qa} and 
tetra-quark interpretations~\cite{Guo:2011gu,Richard:2011} 
using the analogy to the charm sector. Also, in~\cite{Danilkin:2011sh} 
the new resonances were identified as a hadro-quarkonium system 
based on the channel coupling of light and heavy quarkonia to 
intermediate open-flavor heavy-light mesons. 
 
In this paper we analyze the two-body strong decays $\US(nS) \pi^+$ 
of $\zb$ and $\zbp$ using a phenomenological Lagrangian 
approach developed in Refs.~\cite{Faessler:2007gv}-\cite{Dong:2010gu}
which is based on the compositeness 
condition~\cite{Weinberg:1962hj,Efimov:1993ei}. 
In particular, in~\cite{Faessler:2007gv}--\cite{Dong:2010gu}
recently observed unusual hadron states (like $D_{s0}^*(2317)$, 
$D_{s1}(2460)$, $X(3872)$, $Y(3940)$, $Y(4140)$, $Z(4430)$, 
$\Lambda_c(2940)$, 
$\Sigma_c(2800)$) were analyzed within the structure assumption as hadronic 
molecules. The compositeness condition implies that the renormalization  
constant of the hadron wave function is set equal to zero or that the hadron 
exists as a bound state of its constituents. It was originally applied to the 
study of the deuteron as a bound state of proton and 
neutron~\cite{Weinberg:1962hj} (see also Ref.~\cite{Dong:2008mt} 
for a further application 
of this approach to the case of the deuteron). Then it was extensively used 
in low--energy hadron phenomenology as the master equation for the treatment 
of mesons and baryons as bound states of light and heavy constituent quarks 
(see e.g. Refs.~\cite {Efimov:1993ei,Anikin:1995cf,Dubnicka:2010kz}). 
By constructing a phenomenological Lagrangian including the couplings 
of the bound state to its constituents and the constituents to other final 
state particles, we evaluated meson--loop diagrams which describe the 
different decay modes of the molecular states 
(see details in~\cite{Faessler:2007gv}).  

In the present report we proceed as follows. In Sec.~II we briefly review the 
basic ideas of our approach. We now consider the two new resonances
$\zb$ and $\zbp$ as the two molecular states of $\bar{B}B^*$ and 
$\bar{B}^*B^*$. Then we proceed to estimate their strong two-body  
decays $\zb\to \US(nS)+\pi^+$ and $\zbp\to \US(nS)+\pi^+$ where $n=1,2,3$ 
based on an phenomenological Lagrangian approach.
In Sec.~III we present our numerical results and 
a short summary is given in Sec. ~IV.

\section{Phenomenological Lagrangian approach}

Here we briefly discuss the formalism for the study of the composite 
(molecular) structure of the $\zb$ and $\zbp$ resonances. 
In the following calculation we adopt the spin and parity quantum numbers
$J^P = 1^{+}$ for the two resonances $\zb$ and $\zbp$. We consider 
these two new charged hidden-bottom meson resonances as a superposition 
of molecular states of  $\bar{B}B^{*}$ and $\bar{B}^*B^*$ as
\eq
|Z_b^+(10610)\ra &=&\frac{1}{\sqrt{2}} \Big| B^{*+}\bar{B}^0+\bar{B}^{*0}B^+
\Big\ra,
\nonumber \\
|Z_b^{+'}(10650)\ra &=& |B^{*+}\bar{B}^{*0}\ra.
\en 
Our approach is based on an interaction Lagrangian describing 
the coupling of the $\zb$ (or $\zbp$) to its constituents. The simplest 
forms of such  Lagrangians read 
\eq\label{Lagr}
\hspace*{-1cm}
{\cal L}_{Z_b}(x)&=&\frac{g_{_{Z_b}}}{\sqrt{2}}\, M_{Z_b} \, 
Z_b^{\mu}(x)\int d^4y \, \Phi_{Z_b}(y^2)\Big (B(x+y/2)
\bar{B}^*_{\mu}(x-y/2)\nonumber\\
&+&B^*_{\mu}(x+y/2)\bar{B}(x-y/2)\Big ), \\ 
\hspace*{-1cm}
{\cal L}_{Z_b'}(x)&=&\frac{g_{_{Z_b'}}}{\sqrt{2}} \, 
i \epsilon_{\mu\nu\alpha\beta} 
\partial^{\mu}Z_b^{'\nu}(x) \int d^4y \, 
\Phi_{Z_b'}(y^2)B^{*\alpha}(x+y/2)\bar{B}^{*\beta}(x-y/2), 
\en
where $y$ is a relative Jacobi coordinate,  
$g_{_{Z_b}}$ and $g_{_{Z_b^\prime}}$ are the dimensionless 
coupling constants of $\zb$ and $\zbp$ to the molecular $\bar{B}B^{*}$ and 
$\bar{B}^*B^{*}$ 
components, respectively. Here $\Phi_{Z_b}(y^2)$ and $\Phi_{Z_b^\prime}(y^2)$ 
are correlation functions, which describe the distributions of 
the constituent mesons in the bound states. 
A basic requirement for the 
choice of an explicit form of the correlation function $\Phi_H(y^2)$ 
($H = Z_b, Z_b^\prime$) is that 
its Fourier transform vanishes sufficiently fast in the ultraviolet region 
of Euclidean space to render the Feynman diagrams ultraviolet finite. We 
adopt a Gaussian form for the correlation function. The Fourier transform of 
this vertex function is given by 
\eq\label{corr_fun}
\tilde\Phi_H(p_E^2/\Lambda^2) \doteq \exp( - p_E^2/\Lambda^2)\,, 
\en  
where $p_{E}$ is the Euclidean Jacobi momentum.
$\Lambda$ is a size parameter characterizing the distribution of the two 
constituent mesons in the $\zb$ and $\zbp$ systems, which also leads to a 
regularization of the ultraviolet divergences in the Feynman diagrams.  
From our previous analyses of the strong two-body decays of the 
$X, Y, Z$ meson resonances and of the $\Lambda_c(2940)$ and $\Sigma_c(2880)$  
baryon states we deduced a value of $\Lambda \sim 1$~GeV~\cite{Dong:2009tg}.  
For a very loosely bound system like the $X(3872)$ a size parameter of 
$\Lambda \sim 0.5$~GeV~\cite{Dong:2009uf} is more suitable.
The coupling constants $g_{_{Z_b}}$ and $g_{_{Z_b^\prime}}$ are then
determined by the  compositeness condition~\cite{Weinberg:1962hj, 
Efimov:1993ei,Anikin:1995cf, Dong:2009tg,Faessler:2007gv}. It implies 
that the renormalization constant of the hadron wave function is set equal 
to zero with:
\eq\label{ZLc} 
Z_H  = 1 - \Sigma_H^\prime(M_H^2) = 0 \,. 
\en
Here, $\Sigma_H^\prime$ is the derivative of the transverse part of the 
mass operator $\Sigma_H^{\mu\nu}$ of the molecular states (see Fig.1), 
which is defined as 
\eq 
\Sigma_H^{\mu\nu}(p) = g^{\mu\nu}_\perp \, \Sigma_H(p) 
+ \frac{p^\mu p^\nu}{p^2} 
\Sigma_H^L(p)\,, \quad 
g^{\mu\nu}_\perp = g^{\mu\nu} - \frac{p^\mu p^\nu}{p^2} \,. 
\en 
The compositeness condition~(\ref{ZLc}) gives a constraint 
on the choice of the free parameter~$\Lambda$.  
Analytical expressions for the couplings $g_{_{Z_b}}$ and $g_{_{Z_b^\prime}}$ 
are given in Appendix A. 
In the calculation the masses of $Z_b$ and $Z_b^\prime$ 
are expressed in terms of the constituent masses and the binding 
energy $\epsilon$ (a variable quantity in our calculations): 
\eq 
M_{Z_b}=M_{B}+M_{B^*}-\epsilon\,, \quad 
M_{Z_b^\prime}=2M_{B^*}-\epsilon\,. 
\en 
Here we assume bound states for the $Z_b$ and $Z_b^\prime$. 

In the calculation of the two-body decays 
$\zb\to \US(nS)+\pi^+$ and $\zbp\to \US(nS)+\pi^+$ 
we include the direct four-point interactions for the $BB^*\US\pi^+$ and 
$B^*B^*\US\pi^+$ vertices.
The respective phenomenological Lagrangians take the form
\eq
{\cal L}_{BB^*\US\pi}(x)&=&g_{_{BB^*\US\pi}}\US_{\mu}(x)
\bar{B}^{*\mu}(x)\vec{\pi}(x)\cdot\vec{\tau}B(x) + \mathrm{H.c.}\,,
\label{BBSUPpi}\\ 
{\cal L}_{B^*B^*\US\pi}(x)&=&i\epsilon_{\mu\nu\alpha\beta}
\Big (g_{_{B^*B^*\US\pi}}\US^{\mu}(x)
\bar{B}^{*\beta}(x)\partial^{\nu}\vec{\pi}(x)
\cdot\vec{\tau}B^{*\alpha}(x)\nonumber\\ 
&+&f_{_{B^*B^*\US\pi}}\partial^{\nu}\US^{\mu}(x)
\bar{B}^{*\beta}(x)\vec{\pi}(x)
\cdot\vec{\tau}B^{*\alpha}(x)\Big ) \,. \label{BSBSUPpi}
\en 
The four-particle coupling constants defined in Eqs.~(\ref{BBSUPpi}) and 
(\ref{BSBSUPpi}) are effective, which also include off-shell effects. Such couplings 
are obviously different from the strong couplings of molecular states to 
their constituents which model their composite structure via vertex function 
distibutions. We use effective Lagrangians (using both SU(4) and SU(5) 
classification schemes) developed by Ko and collaborators~\cite{Lin:2000ke} 
which worked phenomenologically quite successfully.  
Corresponding diagrams contributing to the $\zb \to \US(nS)+\pi^+$ 
and $\zbp\to \US(nS)+\pi^+$ processes are shown in Fig.2. 

\begin{figure}
\centering
\includegraphics [scale=0.5]{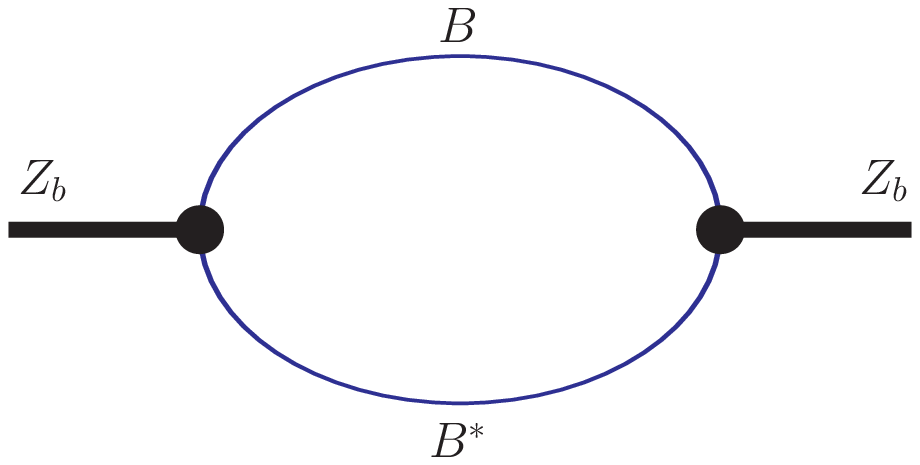}
\hspace{1cm}
\includegraphics [scale=0.5]{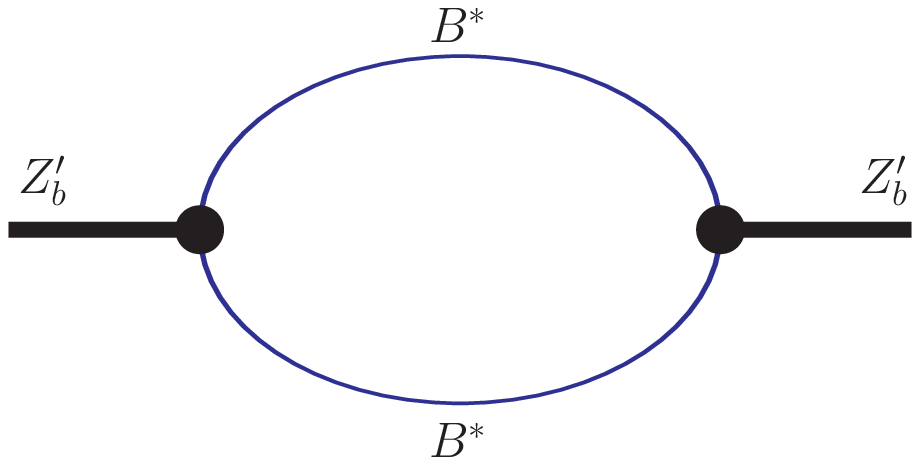}
\caption{Mass operators of $\zb$ and $\zbp$.}
\vspace*{-1.5cm} 
\includegraphics [scale=0.6]{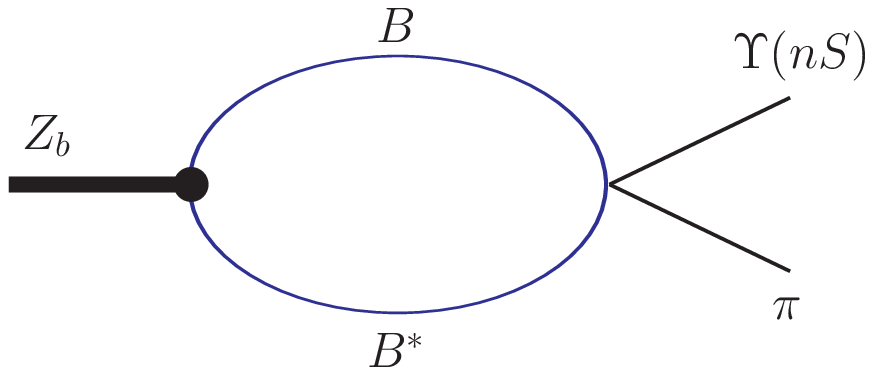}
\hspace*{.5cm}
\includegraphics [scale=0.6]{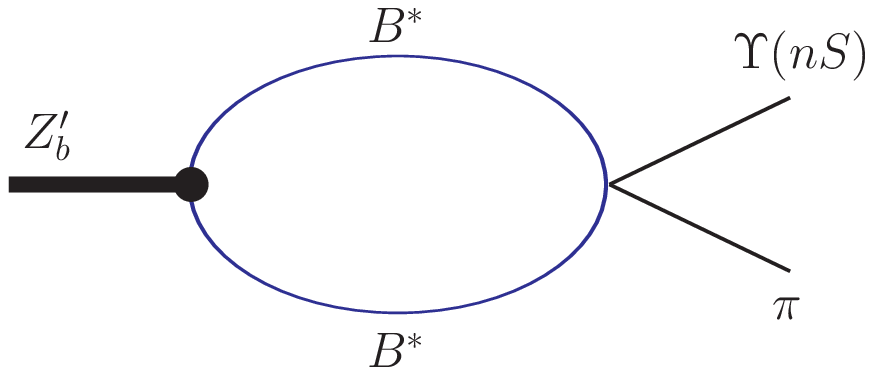}
\vspace*{1.cm}
\caption{Two-body decays $\zb \to \Upsilon(nS) + \pi$ and 
$\zbp \to \Upsilon(nS) + \pi$.} 
\end{figure}

Among the three couplings $g_{_{BB^*\US\pi}}$, 
$g_{_{B^*B^*\US\pi}}$ and $f_{_{B^*B^*\US\pi}}$ we have the 
relations~\cite{Lin:2000ke} 
\eq\label{relations} 
g_{_{BB^*\US\pi}} \ = \ \frac{g_{_{\US BB}} \, g_{_{B^*B\pi}}}{2\sqrt{2}} 
\,,\quad\quad  
g_{_{B^*B^*\US\pi }} \ = \ f_{_{B^*B^*\US\pi}}=\frac{g_{_{BB^*\US\pi}}}
{2\sqrt{M_BM_{B^*}}} \,. 
\en 
The hadronic couplings $g_{_{\US BB}}$ and $g_{_{B^*B\pi}}$ 
are defined as~\cite{Dong:2009yp,Dong:2009uf}: 
\eq 
{\cal L}_{B^\ast B \pi}(x) &=& \frac{g_{_{B^\ast B \pi}}}{\sqrt{2}} \, 
\bar B^\ast_\mu(x)  i\partial^\mu \vec{\pi}(x) \, \vec{\tau} \, 
B(x) + {\rm H.c.}  \,, \nonumber\\
{\cal L}_{\US BB}(x) &=& g_{_{\US BB}} \, \US_\mu(x) 
\bar B(x) i \partial^\mu B(x) + {\rm H.c.} 
\en 
The coupling constant $g_{_{\US(nS)BB}}$ is given by 
\eq\label{V_universality} 
g_{_{\US(nS)BB}}=  \frac{M_{\US(nS)}}{f_{\US(nS)}} \,, 
\en 
where $f_{\US(nS)}$ is 
determined from the leptonic decays of the $\US(nS)$ states as
\eq
\Gamma \Big (\US(nS)\to e^+e^-\Big)=
\frac{4\pi\alpha_{_{\rm EM}}^2}{27}\frac{f^2_{\US(nS)}}
{M_{\US(nS)}}\,, 
\en 
where $\alpha_{_{\rm EM}} = 1/137.036$ is the fine-structure constant. 
The relation (\ref{V_universality}) is analogue of the $\rho$-meson 
universality 
\eq 
g_{_{\rho\pi\pi}} = \frac{M_\rho}{f_\rho} = \frac{1}{g_{\rho\gamma}}
\en  
extended to the bottom sector in Ref.~\cite{Lin:2000ke}, where 
$g_{\rho\gamma}$ is the $\rho\to\gamma$ transition coupling. 
Here
For the last couplings we get 
$f_{\US(1S)}=715.2$ MeV, 
$f_{\US(2S)}=497.5$ MeV, 
$f_{\US(3S)}=430.2$ MeV,
where we used the mass values $M_{\US(1s,2s,3s)}=9460.30\pm 0.26~$MeV, 
$10023.26 \pm 0.31~$MeV and $10355.2 \pm 0.5~$MeV as well as the results 
for the leptonic decay widths of the $\US(nS)$ states 
\eq
\Gamma\Big (\US(1S)\to e^+e^-\Big )&=&1.340\pm 0.018~\mathrm{keV}\,,
\nonumber\\ 
\Gamma\Big (\US(2S)\to e^+e^-\Big )&=&0.612\pm 0.011~\mathrm{keV}\,,
\nonumber\\ 
\Gamma\Big (\US(3S)\to e^+e^-\Big )&=&0.443\pm 0.008~\mathrm{keV}\, .
\en 
Note that we explicitly take into account the $M_{\US(nS)}$ dependence of 
the $f_{\US(nS)}$ and $g_{_{\US BB}}$ couplings. 

The coupling $g_{_{BB^*\pi}}$ can be related 
to the effective coupling constant $\hat{g}=0.44\pm 0.03^{+0.01}_{-0.00}$ 
determined in a lattice calculation~\cite{Becirevic:2009yb}. 
The relation is
\eq
g_{B^*B\pi}=\frac{2\hat{g}}{f_{\pi}}\sqrt{M_BM_{B^*}} \simeq 35.34,
\en
where $f_\pi \simeq 132$ MeV is the pion decay constant. 
We should stress that the phenomenological Lagrangians developed 
in Refs.~\cite{Lin:2000ke} were successfully applied to different 
aspects of heavy flavor physics.

\section{Numerical results}

With the phenomenological Lagrangians introduced and discussed in Sec.~II  
one can proceed to determine the widths of the two-body decays 
$\zb\to \US(ns)+\pi^+$ and $\zbp\to \US(ns)+\pi^+$ with $n=1,2,3$ 
(see corresponding diagrams in Fig.2).  
The corresponding decay widths are given by: 
\eq 
\Gamma_{\zb\to\US(nS)\pi^+}&\simeq& 
 \frac{g_{Z_b\Upsilon(nS)\pi}^2}{16\pi M_{Z_b}}
\lambda^{1/2}(M_{Z_b}^2, M_{\US(nS)}^2, M_{\pi}^2) \,, \nonumber\\
&&\\
\Gamma_{\zbp\to\US(nS)\pi^+}&\simeq&  
\frac{g_{Z_b^\prime\Upsilon(nS)\pi}^2}{16 \pi M_{Z_b^\prime}}
\lambda^{1/2}(M_{Z_b^\prime}^2, M_{\US(nS)}^2, M_{\pi}^2) \,, \nonumber 
\en
where $\lambda(x,y,z)=x^2+y^2+z^2-2xy-2xz-2yz$ 
is the K\"allen function.
The decay coupling constants 
$g_{Z_b\US(nS)\pi}$ and $g_{Z_b^\prime\US(nS)\pi}$ 
involve the products 
$g_{Z_b\US(nS)\pi} = g_{_{Z_b}} \, g_{_{BB^*\US(nS)\pi}} \, J_1$ 
and $g_{Z_b^\prime\US(nS)\pi} = g_{_{Z_b^\prime}} \, 
g_{_{B^*B^*\US(nS)\pi}} \, M_{Z_b^\prime} \, J_2$ where the loop integrals 
$J_1$ and $J_2$ are given in Appendix A.  

For our numerical evaluation hadron masses are taken from the compilation of 
the PDG~\cite{PDG:2010}. The only free parameter of our calculation is the 
dimensional parameter $\Lambda$ entering in the correlation function of 
Eq.~(\ref{corr_fun}). As mentioned before, the parameter 
$\Lambda$ describes the distributions of $BB^*$ and $B^*B^*$ in the $\zb$ 
and $\zbp$ bound state systems, respectively. 
Tables I and II contain
our estimates for the 
decay widths of $\zb\to\US(nS)+\pi^+$ and $\zbp\to\US(nS)+\pi^+$. 
We also indicate the values for the couplings $g_{_{Z_b}}$ and 
$g_{_{Z_b^\prime}}$ as determined from the compositeness condition. 
We find that the data on the strong $\US(nS)\pi$ decays of 
$Z_b$ and $Z_b^\prime$ can be 
qualitatively described for $\Lambda \simeq 0.5$ GeV. This value is close
to the one used for the molecular system of the $X(3872)$\cite{Dong:2009uf}.  
This also means that the $Z_b$ and $Z_b^\prime$ states are considered as 
extended molecular states. 
For transparency we present the results for several choices of the size  
parameter $\Lambda=0.4, 0.45, 0.5, 0.55$~GeV and the binding energy 
$\epsilon$. 
Note that an increase of $\Lambda$ leads to a larger decay width. 
Although the dependence of the decay widths on the binding energy is rather
moderate, a quantitative prediction for these decays strongly depends on 
the size of the system. The decay pattern of $Z_b$ and $Z_b^\prime$ 
to $\US(nS)+\pi^+$, that is the relative importance of the 
$\US(nS)+\pi^+$ decay channels for $n=1,2,3$, can be reproduced and 
could give further support for the molecular interpretation of these states. 
One can see from Tables I and II  
that the rates are increased by a factor 2-2.5 which means that the 
amplitudes are roughly increased by a factor 1.4-1.6. 
Latter value is consistent with a growth of the cutoff parameter of 
$0.55/0.4=1.375$. In this respect the calculation is consistent and 
we do not see any disagreement. It is also clear that any model based 
on cutoff regularization has cutoff-dependent result. 
Here the cutoff is related to the size of hadronic molecular compound. 
As done in our previous analyses of other heavy hadron molecules 
data help to do a fine tuning of the cutoff parameter which is specific 
for a particular molecular state. 
There is no a universal cutoff for the whole tower of possible hadronic 
molecular states. Our previous analyses of molecular states consisting 
of two heavy mesons (like the $X(3872)$ state) indicate that the cutoff 
parameter in the vertex functions of $Z_b$ and $Z_b'$ states should be 
around 0.5 GeV. This is the reason why we do predictions for the valus 
of $\Lambda$ close to 0.5 GeV and are also waiting for more precise data 
with smaller error bars. Please also note that the ratio of rates is 
less dependent on the cutoff and hence a stabile prediction of the model. 
Since data are given for the absolute rates we also choose to use this 
presentation for our results and not the relative rates.

\begin{center}
{\bf Table I.} $Z_b^+\to \US(nS)+\pi^+$ decay properties. 

\vspace*{.15cm} 

\hspace*{-1.25cm}
\begin{tabular}{|c|c|c|c|c|}\hline
$\epsilon(\mathrm{MeV})$ &$g_{Z_b}$ 
&$\Gamma_{1S}(\mathrm{MeV})$ 
&$\Gamma_{2S}(\mathrm{MeV})$ 
&$\Gamma_{3S}(\mathrm{MeV})$\\ \hline
1 &3.3, 3.4, 3.6, 3.7 &11.6, 16.4, 22.3, 29.5  &13.7, 19.4, 26.4, 34.9 &7.2, 10.2, 13.9, 18.4\\ 
\hline
5 &4.0, 4.0, 4.1, 4.2 &11.0, 16.5, 22.4, 29.5  &13.0, 19.4, 26.3, 34.7 &6.7, 10.1, 13.6, 18.0\\ 
\hline
10 &5.0, 4.9, 4.8, 4.8 &11.7, 16.9, 22.7, 29.7 &13.7, 19.8, 26.6, 34.8 &7.0, 10.1, 13.5, 17.6\\ 
\hline
20 &7.2, 6.6, 6.3, 6.0 &13.3, 18.3, 24.0, 30.8 &15.4, 21.2, 27.8, 35.7 &7.4, 10.2, 13.4, 17.3\\ 
\hline
30 &9.4, 8.5, 7.9, 7.4 &14.5, 19.6, 25.4, 32.3 &16.7, 22.5, 29.2, 37.1 &7.7, 10.3, 13.4, 17.0\\ 
\hline
40 &11.7, 10.4, 9.5, 8.8&15.4, 20.4, 26.7, 33.7 &17.5, 23.2, 30.3, 38.3 &7.5, 10.0, 13.1, 16.5\\ 
\hline
50 &13.9, 12.3, 11.1, 10.2&15.9, 21.2, 27.6, 34.9&17.8, 23.9, 31.1, 39.2&7.1, 9.6, 12.5, 15.7\\ 
\hline
Exp. &  &$22.9\pm 7.3\pm 2$ &$21.1\pm 4^{+2}_{-3}$ &$12.2\pm 1.7\pm 4$ 
\\ \hline
\end{tabular}
\end{center}

\vspace*{.15cm} 

\begin{center}
{\bf Table II.} $\zbp \to \US(nS) + \pi^+$ decay properties. 
            
\vspace*{.15cm} 

\hspace*{-1.25cm}
\begin{tabular}{|c|c|c|c|c|c|}\hline
$\epsilon(\mathrm{MeV})$ &$g_{Z_b^\prime}$ 
&$\Gamma_{1S}(\mathrm{MeV})$ 
&$\Gamma_{2S}(\mathrm{MeV})$ 
&$\Gamma_{3S}(\mathrm{MeV})$\\ \hline
1  &3.2, 3.4, 3.6, 3.7 &12.0, 16.9, 23.0, 30.4 &14.7, 20.8, 28.3, 37.4 &9.0, 12.8, 17.4, 23.0\\ 
\hline 
5  &3.9, 4.0, 4.1, 4.2 &12.1, 17.0, 23.0, 30.3 &14.9, 20.8, 28.2, 37.2 &9.0, 12.6, 17.1, 22.6\\ 
\hline
10 &5.0, 4.8, 4.8, 4.7 &12.7, 17.4, 23.4, 30.6 &15.4, 21.3, 28.5, 37.3 &9.2, 12.7, 17.1, 22.3\\ 
\hline
20 &7.2, 6.6, 6.3, 6.0 &14.0, 18.8, 24.6, 31.7 &16.9, 22.7, 29.8, 39.3 &9.8, 13.2, 17.3, 22.3\\ 
\hline
30 &9.4, 8.5, 7.8, 7.3 &15.1, 20.1, 26.1, 33.1 &18.1, 24.1, 31.3, 39.7 &10.2, 13.5, 17.6, 22.3\\ 
\hline
40 &11.7, 10.4, 9.4, 8.7&15.8, 21.1, 27.3, 34.5 &18.9, 25.1, 32.4, 41.0 &10.2, 13.6, 17.6, 22.2\\ 
\hline
50 &13.9, 12.3, 11.0, 10.1&16.2, 21.7, 28.2, 35.6&19.1, 25.5, 33.2, 41.9&9.9, 13.3, 17.3, 21.8\\ 
\hline   
Exp. & &$12\pm 10\pm 3$ &$16.4\pm 3.6^{+4}_{-6}$ &$10.9\pm 2.6^{+4}_{-2}$\\ 
\hline
\end{tabular}
\end{center}

\section{Summary}

To summarize, we have pursued a hadronic molecular interpretation for the 
two recently observed hidden-bottom charged mesons $\zb$ and $\zbp$.
In our calculation we have used the spin-parity assignment $J^P=1^+$ for 
the two resonances, which is currently favored 
by the experimental decay distributions.
We have studied the consequences for their two-body decays 
$\zb \to \US(nS)+\pi^+$ and $\zbp \to \US(nS)+\pi^+$ 
within a phenomenological Lagrangian approach.  
The calculated results for the decay widths (see Tables I and II)  
are of the order of MeV and for the most part 
qualitatively consistent with the numbers deduced by the Belle collaboration.
Especially the experimental results for $\zb(\zbp) \to \US(nS)+\pi^+$ can
be reproduced taking a value for the free size parameter $\Lambda $ 
near 0.5 GeV. 

\section*{Acknowledgments}

This work is supported by the DFG under Contract No. LY 114/2-1, 
National Sciences Foundations of China No.10975146 and 11035006, 
Federal Targeted Program "Scientific and scientific-pedagogical personnel 
of innovative Russia" Contract No.02.740.11.0238. 
One of us (YBD) thanks the Institute of Theoretical Physics, 
University of T\"ubingen for the warm hospitality and thanks the support 
from the Alexander von Humboldt Foundation. 

\appendix 
\section{Coupling constants and structure integrals} 

The expressions for the coupling constants $g_{_{Z_b}}, g_{_{Z_b'}}$ 
and structure integrals $J_1$, $J_2$ are
\eq 
g_{_{Z_b}}^{-2} &=& \frac{M_{Z_b}^2}{32\pi^2\Lambda^2} 
\, \int\limits_0^\infty \frac{d\alpha_1 d\alpha_2}{\Delta_1^3} \, 
(\alpha_{12} + 2 \alpha_1 \alpha_2)    
\left( 1 + \frac{\Lambda^2}{2M_{B^*}^2\Delta_1} \right) \nonumber\\ 
&\times&\exp\left\{ -\frac{M_{B^*}^2\alpha_1 + M_{B}^2 \alpha_2}{\Lambda^2}
+ \frac{M_{Z_b}^2}{2\Lambda^2}  \, 
\frac{\alpha_{12} + 2 \alpha_1 \alpha_2}{\Delta_1}\right\}\,,\\ 
g_{_{Z_b'}}^{-2}&=& \frac{M_{Z_b^\prime}^2}{16\pi^2\Lambda^2} 
\, \int\limits_0^\infty \frac{d\alpha_1 d\alpha_2}{\Delta_1^2} \, 
\left( \frac{\Lambda^2}{M_{Z_b'}^2} 
+ \frac{\alpha_{12} + 2 \alpha_1 \alpha_2}{2\Delta_1} \right) 
\left( 1 + \frac{\Lambda^2}{M_{B^*}^2\Delta_1} \right) \nonumber\\ 
&\times&\exp\left\{ -\frac{M_{B^*}^2\alpha_{12}}{\Lambda^2} 
+ \frac{M_{Z_b'}^2}{2\Lambda^2}  \, 
\frac{\alpha_{12} + 2 \alpha_1 \alpha_2}{\Delta_1}\right\}\,, \\ 
J_1 &=& \frac{1}{8 \pi^2} \, \int\limits_0^\infty \, 
\frac{d\alpha_1 d\alpha_2}{\Delta_2^2} \, 
\left( 1+\frac{\Lambda^2}{2M_{B^*}^2 \Delta_2} \right) \nonumber\\ 
&\times&\exp\left\{ -\frac{M_{B^*}^2\alpha_1 + M_{B}^2 \alpha_2}{\Lambda^2}
+ \frac{M_{Z_b}^2}{4\Lambda^2}  \, 
\frac{\alpha_{12} + 4 \alpha_1 \alpha_2}{\Delta_2}\right\}\,,\\
J_2 &=& \frac{1}{8 \, \pi^2} \, \int\limits_0^\infty \, 
\frac{d\alpha_1 d\alpha_2}{\Delta_2^2} \, 
\left( 1+\frac{\Lambda^2}{M_{B^*}^2 \Delta_2} \right) \nonumber\\ 
&\times&\exp\left\{ -\frac{M_{B^*}^2\alpha_{12}}{\Lambda^2} 
+ \frac{M_{Z_b'}^2}{4\Lambda^2}  \, 
\frac{\alpha_{12} + 4 \alpha_1 \alpha_2}{\Delta_2}\right\}\,,  
\en
where 
\eq 
\Delta_1=2 + \alpha_{12} \,, \quad \Delta_2=1 + \alpha_{12} \,, \quad 
\alpha_{12} = \alpha_1 + \alpha_2 \,. 
\en

\end{document}